\documentclass{article}
\usepackage{ijcai24}

\usepackage{times}
\usepackage{soul}
\usepackage{url}
\usepackage[hidelinks]{hyperref}
\usepackage[utf8]{inputenc}
\usepackage[small]{caption}
\usepackage{graphicx}
\usepackage{amsmath}
\usepackage{amsthm}
\usepackage{booktabs}
\usepackage{algorithm}
\usepackage{algorithmic}
\usepackage[switch]{lineno}

\usepackage{subcaption}
\usepackage{multirow}

\title{Can AI Outperform Human Experts in Creating Social Media Creatives?}

\author{
Eunkyung Park$^1$
\and
Raymond K. Wong$^1$\and
Junbum Kwon$^{1}$
\affiliations
$^1$UNSW Sydney
\emails
\{eunkyung.park, ray.wong,junbum.kwon\}@unsw.edu.au
}

\begin{document}

\maketitle

\begin{abstract}
    Artificial Intelligence (AI) has outperformed human experts in functional tasks such as chess and baduk. How about creative tasks? This paper evaluates AI's capability in {the creative} domain compared to human experts, which little research has been conducted so far. We propose a novel \textbf{Prompt-for-Prompt} to generate social media creatives via prompt augmentation by Large Language Models (LLMs).  
    We take the most popular Instagram posts (with the biggest number of like clicks) in top brands' Instagram accounts to create social media creatives. We give GPT-4 several prompt instructions with text descriptions to generate the most effective prompts for cutting-edge text-to-image generators: Midjourney, DALL-E 3, and Stable Diffusion. LLM-augmented prompts can boost AI's abilities by adding objectives, engagement strategy, lighting and brand consistency for social media image creation. We conduct an extensive human evaluation experiment, and find that (1) AI excels human experts, and (2) Midjourney is better than the other text-to-image generators. Surprisingly, unlike conventional wisdom in the social media industry, prompt instruction including `eye-catching' shows much poorer performance than those including `natural'. Regarding the type of creatives, AI improves creatives with animals or products but less with real people. Also, AI improves creatives with short text descriptions more than with long text descriptions, because there is more room for AI to augment prompts with shorter descriptions.
    
\end{abstract}

\begin{figure}[ht!]
    \begin{subfigure}{\linewidth}
    \includegraphics[scale=0.5]{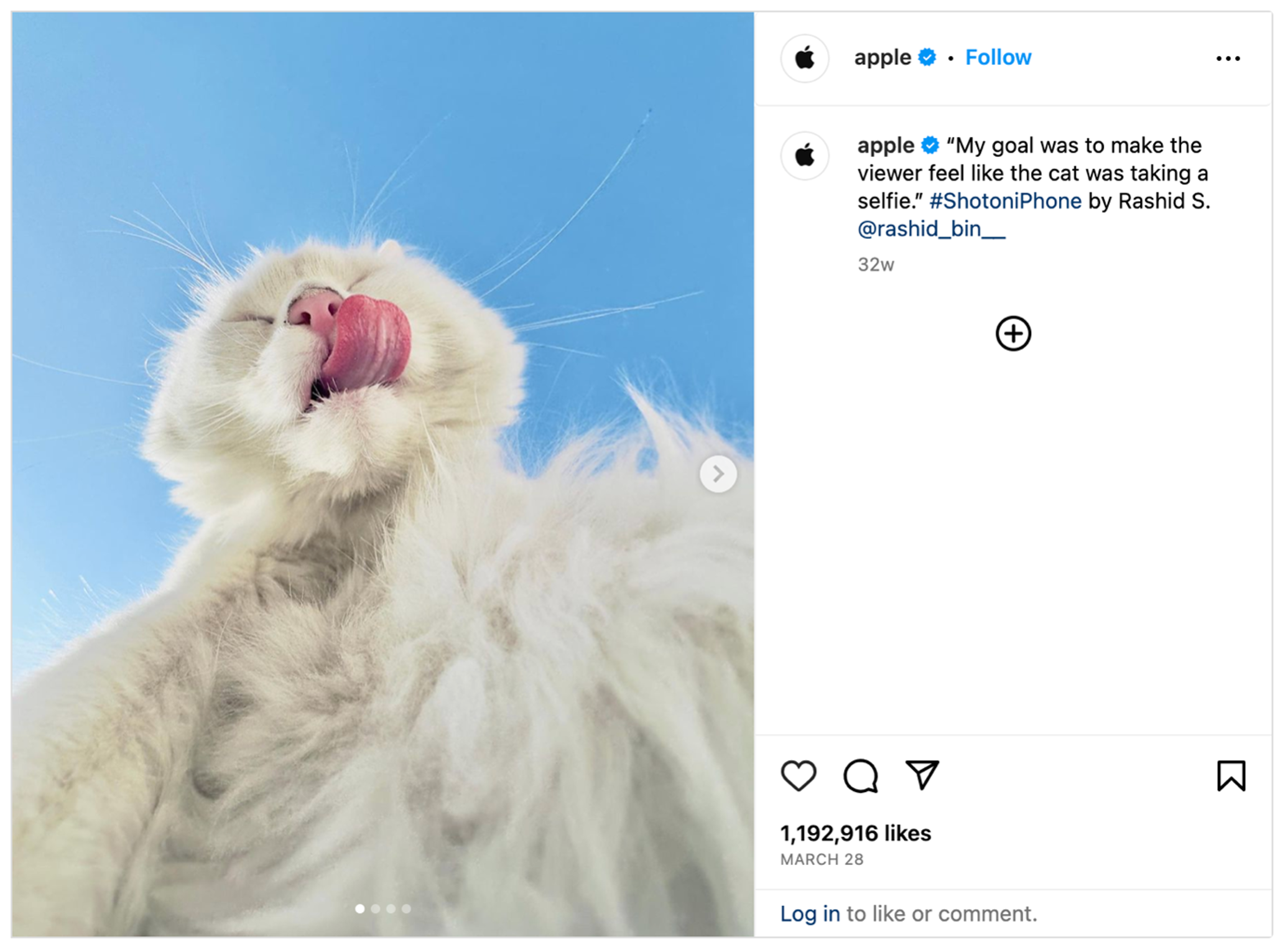}
    \caption{Human-generated social media image from Apple Instagram}
    \label{fig:fig1_a}
    \end{subfigure}
    \begin{subfigure}{\linewidth}
    \includegraphics[scale=.4]{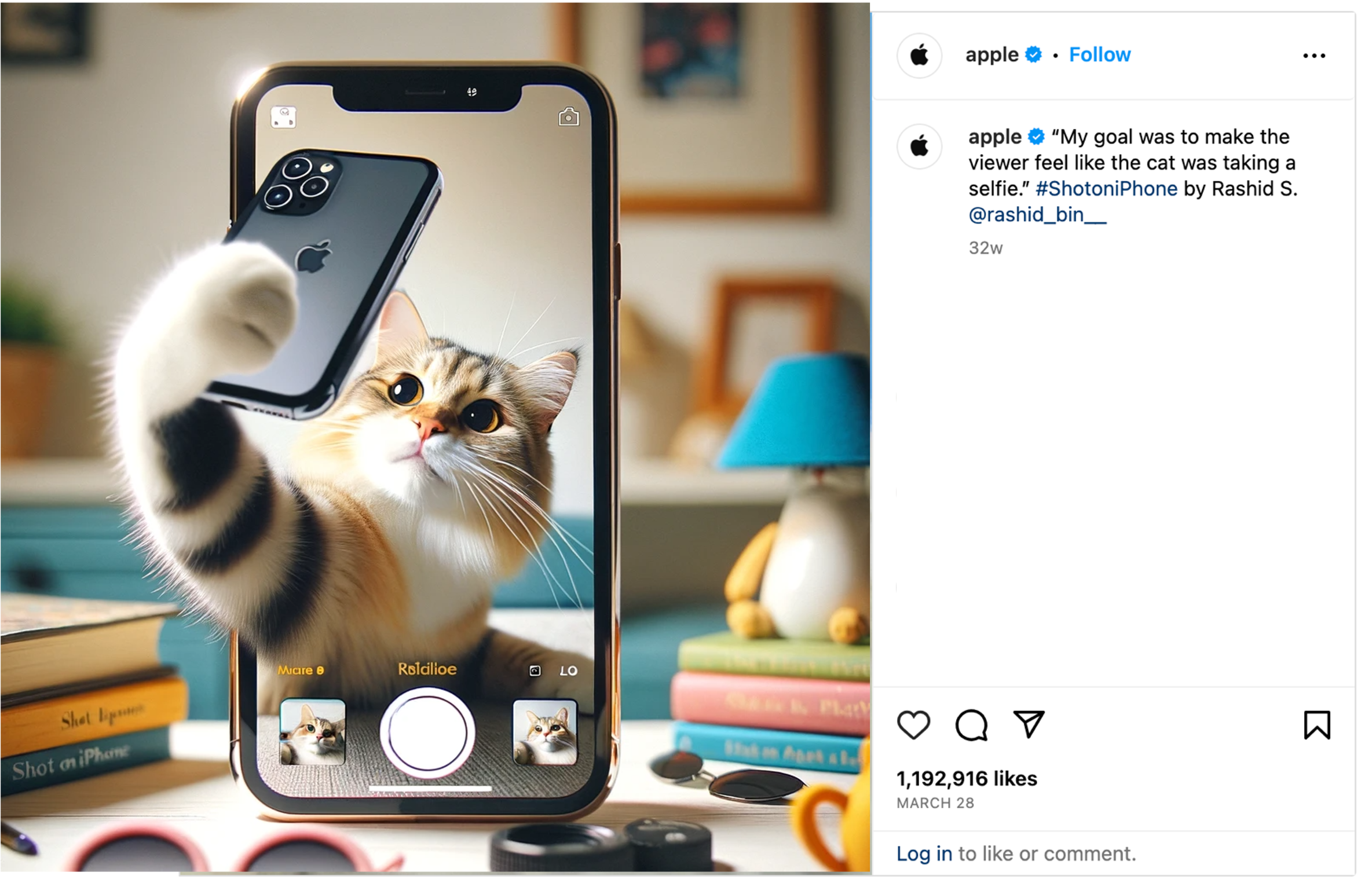}
    \caption{Our \textbf{Prompt-for-Prompt}-generated social media creatives}
    \label{fig1_b}
    \end{subfigure}
    \caption{Human-generated vs AI-generated social media creatives from original text : ``My goal was to make the viewer feel like the cat was taking a selfie."\#ShotoniPhone by Rashid S.@rashid\_bin\_\_}\label{fig:fig1}
\end{figure}

\section{Introduction}
As generative AI products such as ChatGPT become popular, people are growingly worried about whether AI will replace humans across many sectors, including creative domains. For example, Hollywood workers recently had a five-month strike because they were concerned about losing their jobs due to AI \cite{news1}. Creative agencies have also started reducing the number of junior workers because AI can make a draft quickly \cite{news2}. However, while AI has outperformed human experts in functional tasks such as chess and baduk, it has rarely been  documented whether AI is indeed doing a better job than human experts in \textbf{creative tasks}. Leading companies often hire top design experts to make attractive social media creatives. This paper evaluates AI's capability in the creative domain compared to human experts.  
 
Designing an automatic system of social media creatives from textural descriptions, i.e., the text-to-image tasks, is challenging. However, with the advance of deep learning, researchers have shown a substantial improvement in the image generation task. \cite{pmlr-v139-ramesh21a,yu2022scaling} employ large scale data for text-to-image task including DALL-E \cite{pmlr-v139-ramesh21a}. The Diffusion Model (DM) has recently become the new state-of-the-art model in text-to-image generation tasks such as Stable Diffusion \cite{podell2023sdxl}. While the existing research in the text-to-image literature focuses on improving the ability to generate images from ``given" text input, we change the research direction by improving the ``quality" of text input via Large Language Models (LLMs). 
 
The recent success of LLMs has led to growing interest and
prevalence in improving their capability to perform tasks via prompting.
Researchers have published their work regarding
AI's power
compared to human ability. Marked
Personas
\cite{cheng-etal-2023-marked} compares LLMs' outputs with human-written works. A series of research on In-Context Learning
(ICL) such as \cite{chen-etal-2023-many} and CoT
\cite
{NEURIPS2022_9d560961}
also compare AI with humans. Most of the
above research in the LLM literature focuses
on text output or reasoning tasks.   
 
This paper makes the
following contributions and
new findings. First, our research evaluates AI's
ability in Creative tasks compared to human experts and shows that AI-generated
creatives receive higher engagement (i.e., willingness to like and comment) than
creatives of human experts. Our result can give more confidence
to companies in creative industries about adopting generative AI.   
Second, more
importantly, we propose a novel \textbf{Prompt-for-Prompt}
approach to generating social media creatives via prompt augmentation by leveraging
the power of LLMs. Specifically, we take the most
popular Instagram posts (with the biggest number of like clicks) in top brands'
Instagram accounts. Then, we give GPT-4 several prompt instructions with its
text description of such top posts to generate the most effective prompts for cutting-edge
text-to-image generators: Midjourney,
DALL-E, and Stable Diffusion. 
 
For
example, Figure~\ref{fig:fig1}
shows a social media post from Apple Instagram. Figure~\ref{fig:fig1_a}
is the most popular post (made by human experts) with the biggest number of likes within top brands' Instagram accounts. 

Figure~\ref{fig1_b} is an AI-generated image via our prompt
augmentation using the original text description.  
From
the experiments, we find that our \textbf{Prompt-for-Prompt} can boost AI's
abilities by adding creatives’ objectives,
engagement strategy, lighting
and brand consistency for social
media image creation. Our proposed approach is efficient regarding
computation time and cost,
and
can be adapted to many other applications easily beyond social media image
generation. Also, our results show that a
Midjourney
is better than the other text-to-image generators.
 
Third,
we identify effective \textbf{Prompt-for-Prompt} to generate social media creatives.
Surprisingly,
unlike conventional wisdom in the social media industry, prompt instructions 
including ‘eye-catching’ show much poorer performance than that including
`natural’.  
Fourth, we also identify the type of content on which AI (vs human experts) is better or worse. AI improves creatives with animals or products rather than with real people. Also, AI improves creatives with short text descriptions more than with long text descriptions perhaps because there is more room for AI to augment prompts with short descriptions. 

\section{Related Work}

\subsection{Large Language Models (LLMs)}
As LLMs receive great attention from academics and industries, many researchers have tried to uncover the real effects of LLMs outputs \cite{dinan-etal-2020-multi,weidinger2021ethical}. They compare AI-generated outputs to the human-generated outputs. For example, 
Marked Personas \cite{cheng-etal-2023-marked} compares GPT-3.5 and GPT-4 outputs with human-written texts.

In addition, the recent success of Large Language Models has led to growing interest in improving their capability to perform tasks via prompting \cite{NEURIPS2020_1457c0d6,10.1145/3560815}. Some researchers suggested significantly improving LLMs' capability by augmenting the task instructions \cite{10.5555/3455716.3455856,wei2022emergent}. Other lines of research work such as zero-shot CoT \cite{kojima2023large}, auto-CoT \cite{zhang2022automatic}, and Tree of Thoughts (ToT) \cite{yao2023tree} have adapted intermediate language steps since the publication of Chain-of-Thoughts (CoT) \cite{NEURIPS2022_9d560961}, which augments each demo with a chain of rational steps. \cite{chen-etal-2023-many} have studied a demo of In-Context Learning.

Similar to the above studies, our research also uses prompting. However, most of the previous work mainly covers textual data-based tasks. In this paper, we particularly focus on the AI's creativity in text-to-image task.


\begin{table}[t]
    \centering
    \setlength{\tabcolsep}{4pt}
    \begin{tabular}{lll}
        \toprule
           & Prompt instruction &Output \\
        \midrule
         $PI_1$          &  Natural+Social Media Images+ $P0$&$P_{a1}$        \\
        $PI_2$& Eye-Catching+Social Media Images+ $P0$ &$P_{a2}$       \\
                  $PI_3$          & Social Media Images + $P0$&$P_{a3}$       \\
        \bottomrule
    \end{tabular}
    \caption{Examples of prompt instruction. $P0$ is an original text description}
    \label{tab:prompts}
\end{table}

\begin{figure*}
\centering
\includegraphics[scale=0.5]{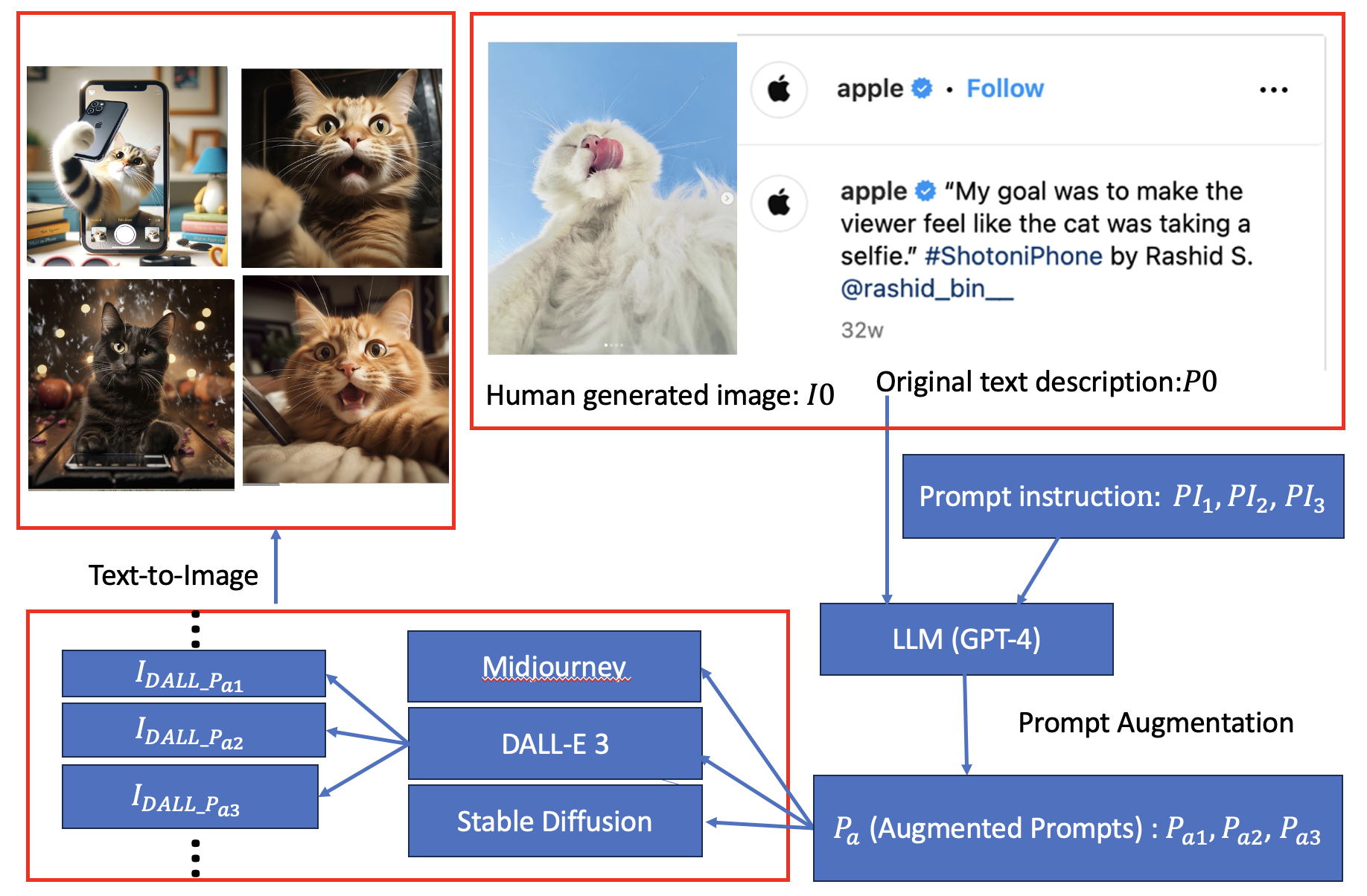}
\caption{Our \textbf{Prompt-for-Prompt}}
\label{fig:inst_apple1}
\end{figure*}

\subsection{Text-to-Image}
Text-to-image is to generate a realistic image from natural language descriptions. 
AlignDRAW \cite{DBLP:journals/corr/MansimovPBS15} is a pioneering work that generates images from natural languages but suffers from unrealistic
results.
\cite{pmlr-v48-reed16} is one of the first to tackle this with a conditional
GAN. From the previous work, multi-stage model \cite{Zhang2016StackGANTT}, attention mechanisms \cite{DBLP:conf/cvpr/XuZHZGH018}, contrastive methods \cite{Zhou_2022_CVPR,DBLP:conf/cvpr/0010KBLY21} have been adapted. \cite{DBLP:conf/icml/RameshPGGVRCS21,DBLP:conf/nips/DingYHZZYLZSYT21} employ sequence modeling approaches. \cite{DBLP:conf/cvpr/EsserRO21,DBLP:journals/tmlr/YuXKLBWVKYAHHPLZBW22} introduce stronger image quantizers and scaling up model parameters. 

Recently \cite{DBLP:conf/icml/NicholDRSMMSC22,DBLP:journals/corr/abs-2204-06125,DBLP:conf/cvpr/RombachBLEO22} apply Diffusion
Models \cite{DBLP:conf/nips/HoJA20} to improve generated image quality. \cite{DBLP:conf/nips/SahariaCSLWDGLA22,DBLP:journals/tmlr/YuXKLBWVKYAHHPLZBW22} scale up text encoder models to achieve
significant gains in generating relevant images.
DALL-E 3 \cite{shi2020improving}  is a text-to-image model by OpenAI, released in Oct 2023. DALL-E 3 is trained from the highly descriptive generated image captions, while existing text-to-image models have difficulty following detailed image descriptions. It uses a bespoke image captioner to recaption the training dataset.

Stable Diffusion XL \cite{podell2023sdxl} is a latent diffusion model for text-to-image synthesis. It applies a three times larger UNet
backbone than previous models and utilises many more model parameters to have more attention
blocks and a larger cross-attention context. Midjourney V5.2 generates images from text description and released in June 2023. 

This paper uses state-of-the-art models such as Midjourney, DALL-E 3, and Stable Diffusion. However, we are not attempting to improve the text-to-image generators by paying computing costs and increasing the parameters. Instead, we propose a novel \textbf{Prompt-for-Prompt} to generate social media creatives via prompt augmentation by LLM (GPT-4). 

\begin{table*}
    \centering
    \begin{tabular}{p{0.3\linewidth} | p{0.3\linewidth}|p{0.3\linewidth}}
        \toprule
        $P_{a1}$&$P_{a2}$&$P_{a3}$ \\ \hline

        \midrule
        \textbf{Objective}: Produce an image that encapsulates the charming and playful moment as if a cat is taking a selfie, capturing the creativity and quality of iPhone photography. \textbf{Engagement Strategy}:Craft the caption or image in a way that encourages viewers to share their own pet selfies taken on an iPhone, possibly creating a user-generated content stream that broadens the campaign's reach. \textbf{Lighting}:Natural light should be utilized to give life to the cat’s fur and eyes, enhancing the image's overall warmth and appeal. \textbf{Props and Accessories}:Include iPhone accessories or props that are typically associated with selfies, like a ring light or a prop to stabilize the iPhone, which can add to the storytelling aspect. &\textbf{Brand Consistency}:Keep the image sleek and consistent with Apple's branding. The composition should be minimalistic yet striking, with a focus on the clever setup that makes the cat appear to take a selfie. \textbf{Call to Action}: Incorporate a subtle call to action, such as “Show us your best selfie \#ShotoniPhone” to encourage user engagement and participation in the campaign. \textbf{Scene Setup}:Position an iPhone in a scene in such a way that a cat appears to be looking into the front camera or tapping the shutter button. This could be done by placing the iPhone against a backdrop where a cat typically lounges, like a sunny windowsill, a cozy corner with scatter cushions, or an interesting outdoor locale. &\textbf{Scene Concept}:Set up a whimsical yet plausible scenario where a cat appears to be using an iPhone to take a selfie. Consider staging the iPhone in a position that could naturally be operated by the cat, such as propped against a pile of books or resting on a soft surface within the cat's reach. \textbf{Cat's Interaction with the iPhone}:Pose the cat with one paw extended toward the iPhone as if pressing the shutter button. Ensure the cat's face is prominently reflected on the iPhone screen, capturing a clear selfie expression. \textbf{Background Setting}:Choose a simple yet visually appealing background that doesn't distract from the subject, such as a brightly lit room with minimal furniture or an outdoor setting with natural light.\\

        \bottomrule
    \end{tabular}
    \caption{Examples of augmented prompts from GPT-4 for social media creatives. Due to the page limitation, we present only part of each augmented prompt.}
    \label{tab:prompts_exp}
\end{table*}

\section{Approach : Prompt-for-Prompt}

Figure~\ref{fig:inst_apple1} shows our overall framework \textbf{Prompt-for-Prompt} for social media image creation.
On the top, there is a social media image and text description. Let $I0$ be the human-generated original image, and $P0$ denotes the original text description in the Instagram post. Therefore, $P0$ is ``apple ``My goal was to make the viewer feel like the cat was taking a selfie" \#ShotoniPhone by Rashid S. @rashid\_bin\_"

\subsection{Prompt Instruction }

From the original text description $P0$, we leverage LLM to generate an augmented prompt. We use 3 prompt instructions ($PI_{1}$, $PI_{2}$, $PI_{3}$) in Table~\ref{tab:prompts} and original text description ($P0$) together as an input for GPT-4. For instance, to make an augmented prompt by the instruction - `eye-catching' and `social media images' ($P_{a2}$), we ask LLM with Prompt instruction ($PI_{2}$), ``Can you create a prompt to make eye-catching social media images for the following description:($P0$)?" by concatenating `eye-catching social media images' and $P0$. Then, LLM creates the augmented prompt $P_{a2}$. Eye-catching content is preferred in social media creatives to increase engagement \cite{info:doi/10.2196/24564}. Also, we select `natural' as another instruction because we worry that AI-generated creatives are not natural from our preliminary experiment.

\subsection{Prompt Augmentation by LLM for Social Media Creatives}

From the text description augmentation using GPT-4, we obtain augmented prompt $P_{a1}$ from the prompt instruction such as $P_{a1}$,$P_{a2}$, $P_{a3}$. Table~\ref{tab:prompts_exp} shows the examples of augmented prompts from GPT-4. It includes key strategies for successful social media image creation, including objectives and Engagement Strategy ($P_{a1}$) to create a natural social media image. It also includes Lighting (Natural light) to create natural social media images. $P_{a2}$ presents a strategy regarding Brand Consistency and Call to Action. It also suggests the Scene Setup concept to create eye-catching social media images. $P_{a3}$ represents a strategy for social media creatives by introducing Cat's interaction and background setting.

Compared to the original text description, which includes only a description of images, augmented prompts suggest critical strategies for social media creatives. Figure~\ref{fig1_b} is generated from our proposed \textbf{Prompt-for-Prompt}. It shows a more detailed cat for taking a selfie by adding a frame of an iPhone and a pose for taking pictures by herself. This improvement is possible due to the augmented prompts' detailed direction, which is not presented in the original text description ($P0$).

We use state-of-the-art text-to-image models such as Midjourney, DALL-E 3, and Stable Diffusion to create social media creatives using the augmented prompts ($P_{a1}$, $P_{a2}$, $P_{a3}$) from the previous step. Let $m$ be a text-to-image method. $m \in \{\text{Midjourney}, \text{DALL-E 3}, \text{Stable Diffusion}\}$. Therefore, we generate social media images $I_{miP_{aj}}$. 


\subsection{Robustness Check and Evaluation Measures}
For qualitative evaluation from human participants, we ask them two questions -  willingness to like and comment for robustness check. The usage of willingness to like and comment on Instagram posts is inspired by recent research \cite{doi:10.1016/j.intmar.2019.06.003}. Commenting requires higher levels of cognitive and behavioural
effort than liking. Thus, writing comments can be considered more
`profound engagement' than clicking the like button \cite{doi:10.1080/00913367.2017.1405753}. Our experiments also find a lower willingness to comment than a willingness to like.

In addition to the two measures, we add further measurements such as Share, Appealing, Relevant to text description, Creative, and Natural to assess social media creatives as image quality and text-image alignment are the two main criteria of text-to-image are \cite{zhang2023texttoimage}. 

\section{Experiments}

\begin{table*}
    \centering
    \begin{tabular}{llrrlrr}
        \toprule
        Brand  & Industry & \# Likes & \# Followers & Image type & Desc len & \# Tasks \\
        \midrule
        Apple & IT & 1193014 & 31.9M & animal-only & 118& 10 \\
        Coca-Cola & Food \& Beverage & 41453 & 3M & product-only & 127 & 10 \\
        Google & IT & 92156 & 14.6M & product-only & 159 & 10 \\
        Herm\'es & Luxury & 126411 & 13.5M & people+product & 250 & 10 \\
        Louis Vuitton & Luxury & 8354713 & 54.8M & people+product & 328	& 10 \\
        Mercedes-Benz&	Car&	1275283&	37.9M&	people+product	&165	&10 \\
        Nike&	Fashion&	2069991&	306M&	people+product&	27&	10 \\
        Rolex&	Luxury&	402222&	15.4M&	product-only&	511&	10 \\
        Starbucks &	Food \& Beverage &	618791 &	18M&	product-only &	206	& 10 \\
        Tesla&	Car&	1041648&	9.4M&	product-only&	72&	10\\
        \bottomrule
    \end{tabular}
    \caption{Statistics of datasets. We collect data from Instagram. }
    \label{tab:dataset}
\end{table*}

\subsection{Tasks and Datasets}

We conduct experiments on 10 Instagram accounts such as Apple, Google, Louis Vuitton, and Tesla from various industries: IT, Fashion, Car, Food \& Beverage. The overall statistics are listed in Table~\ref{tab:dataset}. We collect social media data using the automatic data scraping platform, PhantomBuster (\url{https://phantombuster.com/}). We select top-ranked posts for each brand based on the number of likes. We use social media creatives as human expert-generated creatives. We use text descriptions from the post as input to generate social media creatives with Prompt instruction and text-to-image methods. We rely on a LLM (GPT-4) to generate prompts. For each brand, we generate 10 tasks.
%

\subsection{Text-to-Image Methods and LLM for Prompts}
We use state-of-art text-to-image methods and a LLM.

\begin{itemize}
    \item Midjourney V.5.2 generates images from text descriptions. It is released in June 2023. It combines CLIP (Contrast Language Image Pretraining)
with a constantly
changing set of image generation
methods. One can use the Midjourney Bot on Discord.
    \item DALL-E 3 is a text-to-image model by OpenAI and released in October 2023. To overcome limitations from the training of noisy and inaccurate image captions, it trains a bespoke image captioner to recaption the
training dataset.
    \item Stable Diffusion SDXL is a latent diffusion model for text-to-image and is released in 2023. It uses more attention
blocks and a larger cross-attention context. It introduces a refinement model and post-hoc text-to-image technique.
\end{itemize}

To generate prompts for social media creatives, we select GPT-4. From our preliminary experiments, Meta's Llama 2 (\url{https://www.llama2.ai/}) gives us outputs, including multiple images with short prompts. The outputs from Google's BARD (\url{https://bard.google.com}) are similar to GPT-4. Therefore, we conduct further experiments using only GPT-4 to simplify the experiment.

\subsection{Human Evaluation}

We conduct human surveys using Amazon Mechanical Turk (\url{https://www.mturk.com}). To minimise bias from the ordering effect, we randomise the order of the questionnaire using an online survey tool, Qualtrics (\url{https://www.qualtrics.com/}). We have 100 sets of questionnaires and collect 500 human participants. Human participants are given 10 social media posts, including human-generated creative ($I0$) + text description ($P0$) or AI-generated creative + text description($P0$). Participants are asked questions, including willingness to like and comment. We collect all responses from the survey and calculate the mean values. We model regression models for each question to assess the relationship between our prompt approach to generate social media creatives and target variables (willingness to like/comment), including statistical significance.

\section{Can AI Outperform Human Experts in Creating Social Media Creatives?}



        



To more precisely quantify whether AI-generated creatives from augmented prompts ($P_a$) obtain higher willingness to like/comment, we model the outcome variables received by Instagram's ranked post as a function of $P_a$ where the $P_a$ is computed at the aggregate level by combining
all types of augmented Prompts. In other words, $P_a$=1 if the image is generated from \textbf{Prompt-for-Prompt} or 0 otherwise. 

$\beta_{1}$ measures the difference between human-generated ($I0$) and AI-generated creatives from \textbf{Prompt-for-Prompt} ($P_{a1}$ or $P_{a2}$ or $P_{a3}$). 

The regression model also includes several control variables, such as brand, age, gender, and duration to control other effects: %

\begin{align}
    Like \; or \; Comment = &\beta_{0}+\beta_{1}P_a+\beta_{2}Male \nonumber \\
    + & \beta_{3}Age 
    + \beta_{4}Duration\nonumber \\
    +&\mu_{i}+\tau_{j} + \epsilon
    \label{eq:reg1}
\end{align}%
where : 
\begin{itemize}
\item $\beta_1$ is a core coefficient of interest in this study, which is the main effect of \textbf{Prompt-for-Prompt} (vs human);

    \item $\beta_0$ is a constant term;

     \item $\beta_2$, $\beta_3$, $\beta_4$ capture the effect of male (vs female), Age, and Duration;
     \item $P_a$ is 1 if the image is generated by \textbf{Prompt-for-Prompt} or 0 otherwise;
     
     \item Male is 1 if the respondent is male or 0 otherwise;
     \item Age is an age of a participant;
     \item Duration is a time for completing human survey;
     \item $\mu_{i}$ is dummies for each brand (e.g., Apple, Coca-Cola);
     \item $\tau_{j}$ is dummies for the participants;
     \item $\epsilon$ is an error term.
\end{itemize}

\begin{table}[ht!]
    \centering
    \begin{tabular}{llrrl}
        \toprule
         Response  & Variable&{Coef} & \multicolumn{2}{c}{$p$-value}  \\
         \midrule
         Like&$P_a$&2.48&0.002&***	\\
          
          & Male &-2.38 &1.65e-14  & ****\\
          & Age &0.34 & 0.076  & \textbf{*}\\
          & Duration &0.01 & 6.73e-05 &**** \\
          &\multicolumn{4}{l}{Adjusted R-squared: 0.676} \\
         \hline
         Comment& $P_a$&2.17&0.012&** \\
        & Male &-2.99 &0.009  &*** \\
          & Age & 0.29&0.172 & \\
          & Duration &0.01 & 0.007 &*** \\
          &\multicolumn{4}{l}{Adjusted R-squared: 0.673} \\
        \bottomrule

        \multicolumn{5}{l}{\small *p $<$ 0.1; **p $<$ 0.05; ***p $<$ 0.01;
****p $<$ 0.001.} \\
        
    \end{tabular}
    \caption{Regression result of Model~\ref{eq:reg1}}
    \label{tab:exp1_2}
\end{table}

Table~\ref{tab:exp1_2} shows that people like AI-generated social media creatives more than human expert-generated ones. The key coefficients of interest ($P_{a}$) for both like and comment are positive and significant, respectively. On average, people like creatives from our proposed approach 2.48 (out of 10) more than those from human experts. The model's small $p$-values show the statistically significant differences between the baseline (human expert-generated creatives) and the AI-generated creatives. Similarly, regarding comments, our proposed augmented prompt approach ($P_a$) achieves a 2.17 (out of 10) more willingness to comment compared to the baseline (human-generated creatives), and the difference is statistically significant.
Also, across willingness to like and comment, female (vs male) and older participants give higher willingness to like or comment.

\begin{table}[t!]
    \centering
    \begin{tabular}{llrrl}
        \toprule
         Response  & Variable&{Coef} & \multicolumn{2}{c}{$p$-value}  \\
         \midrule
         Like& $P_{a1}$&	2.49&	0.005& *** \\
         & $P_{a2}$&	1.85&	0.037& ** 	\\ 
         & $P_{a3}$	&2.58&	0.003& *** 	\\
         & Male & -2.34&2.96e-16   & ****\\
          & Age & 0.37& 0.038   & **\\
          & Duration & 0.01& 8.79e-06  &**** \\
          &\multicolumn{4}{l}{Adjusted R-squared: 0.665 } \\ \hline
         Comment& $P_{a1}$&	1.632&	0.090& *   \\
         & $P_{a2}$&	1.856&	0.054& * 	  	\\ 
         & $P_{a3}$	&2.634&	0.006& ***	 	\\
         & Male &-3.11 &-10.127  $<$ 2e-16   & ****\\
          & Age &0.34 & 0.072   & *\\
          & Duration &0.01 & 0.002  &*** \\
          &\multicolumn{4}{l}{Adjusted R-squared: 0.678 } \\
        \bottomrule
        \multicolumn{5}{l}{\small *p $<$ 0.1; **p $<$ 0.05; ***p $<$ 0.01;
****p $<$ 0.001.} \\

    \end{tabular}
    \caption{Regression result from each augmented prompt}
    \label{tab:exp2_2}
    \end{table}

\begin{table}[ht!]
    \centering
    \begin{tabular}{llrrl}
        \toprule
         Response  & Variable&{Coef} & \multicolumn{2}{c}{$p$-value}  \\
         \midrule
         
         Like& Mid+$P_{a1}$&3.02	&	0.006  & *** \\
         & Mid+$P_{a2}$&	 0.97&	0.374&   	\\ 
         & Mid+ $P_{a3}$	&3.57&	0.001 & *** 	\\
         & DALL+$P_{a1}$& 2.81	&	0.010     & ** \\
         & DALL+$P_{a2}$&	 2.37  &	0.030   &**  \\ 
         & DALL+ $P_{a3}$	&2.55&	0.019 & **	\\
         & SD+$P_{a1}$& 1.66	&	0.127     &  \\
         & SD+$P_{a2}$&	 2.24  &	0.040   &**  \\ 
         & SD+ $P_{a3}$	& 1.63&	0.135 & 	\\

         & Male & -2.34&2.96e-16   & ****\\
          & Age & 0.37& 0.038   & **\\
          & Duration & 0.01& 8.79e-06  &**** \\
          &\multicolumn{4}{l}{Adjusted R-squared: 0.665 } \\ \hline
         
         Comment& Mid+$P_{a1}$&1.46	&	0.218      &  \\
         & Mid+$P_{a2}$&	 2.57&	0.029   &**   	\\ 
         & Mid+ $P_{a3}$	& 3.29&	0.005 & *** 	\\
         & DALL+$P_{a1}$& 1.31	&	0.267     &  \\
         & DALL+$P_{a2}$&	 0.67  &	0.570   &  \\ 
         & DALL+ $P_{a3}$	& 2.66&	0.025 & **	\\
         & SD+$P_{a1}$&  2.13	&	0.071429        & * \\
         & SD+$P_{a2}$&	 2.33  &	0.049     &**  \\ 
         & SD+ $P_{a3}$	& 1.96&	0.098    & *	\\

         & Male & -3.11&-10.121  $<$ 2e-16   & ****\\
          & Age &  0.34& 0.072    & *  \\
          & Duration &  0.01& 0.002  &*** \\
          &\multicolumn{4}{l}{Adjusted R-squared: 0.671 } \\ 
        \bottomrule
        \multicolumn{5}{l}{\small *p $<$ 0.1; **p $<$ 0.05; ***p $<$ 0.01;
****p $<$ 0.001.} \\

    \end{tabular}
    \caption{Regression result for text-to-image methods+augmented prompts with human-generated creatives baseline}
    \label{tab:exp2_3}
    \end{table}

\begin{table}[ht!]
    \centering
    \begin{tabular}{l|rrr|rrr}
        \toprule
         & \multicolumn{3}{c}{Like}&\multicolumn{3}{c}{Comment}  \\
         Prompt& Mid&DALL&SD&Mid	&DALL&SD \\
         \midrule
         $P_{a1}$&\textbf{7.14}&6.93&5.78&\textbf{3.99}&3.85&4.67\\
         $P_{a2}$&5.09&\textbf{6.49}&6.36&\textbf{5.11}&3.21&4.87\\
         $P_{a3}$&\textbf{7.69}&6.66&5.75&\textbf{5.83}&5.20&4.50\\
        
        \bottomrule

        \multicolumn{7}{l}{\small Note: Mid-Midjourney, DALL-DALL-E 3, SD-Stable Diffusion} \\

    \end{tabular}
    \caption{Text-to-Image methods comparison}
    \label{tab:tool_comparision}
\end{table}

\begin{table*}[ht!]
    \centering
    \begin{tabular}{llrrrrrrr}
        \toprule
         &&Like&	Comment&	Share&Appealing&	Relevant&Creative&Natural  \\
         \midrule
         \multirow{2}{*}{Product-only} &Human&5.33&1.06&3.74&7.54&7.47&7.47&7.50\\
                &Prompt-AI&6.33&3.36&4.09&7.59&	7.44&	7.47&7.57\\
        \multirow{2}{*}{People+product}&Human&7.54&4.65&	6.82&7.56&	7.26&	7.53&7.53\\
        &Prompt-AI&7.16&5.16&6.52&7.62&7.36&7.51&7.50\\
        \multirow{2}{*}{Animal-only}&Human&7.09&4.36&5.88&
        7.64&7.53&7.36&7.40\\
        &Prompt-AI&8.05&6.04&6.36&7.85&7.55&7.64&	7.62\\

        \bottomrule

    \end{tabular}
    \caption{Human vs AI based on image characteristics}
    \label{tab:char_img}
\end{table*}

\begin{table*}
    \centering
    \begin{tabular}{llrrrrrrrrrr}
        \toprule
         &&Like&	Comment&	Share&Appealing&	Relevant&Creative&Natural  \\
         \midrule
         \multirow{2}{*}{Short} &Human&4.65	&1.87&4.48&		7.65&7.45&7.51&7.65\\
                &Prompt-AI&5.89	&4.01	&4.49	&7.72&7.49&7.53&	7.62	\\
        \multirow{2}{*}{Medium}&Human&7.65&		3.91&5.75&7.48&7.47&7.50&7.37\\
        &Prompt-AI&7.99&5.56&6.05&	7.63	&7.43&7.53&7.47\\
        \multirow{2}{*}{Long}&Human&7.45&		3.01&5.55&7.52&7.24&7.43&7.44\\
        &Prompt-AI&6.92&3.57&5.59&	7.57	&7.30&7.44&7.52\\

        \bottomrule

    \end{tabular}
    \caption{Prompt-for-Prompt effects based on text description length}
    \label{tab:text_leng}
\end{table*}

\section{Ablation Study}

To examine whether each augmented prompt can generate better social media creatives than the creatives by human experts, we model the willingness to like and comment on which Instagram top-ranked posts received as a function of $P_{a1}$, $P_{a2}$, and $P_{a3}$, which are dummy variables. For example, $P_{a1}$=1 when the image is generated from the augmented prompt $P_{a1}$ or 0 otherwise. The
function includes several control variables like the previous regression model. (Refer to the model in Appendix A.1).

Table~\ref{tab:exp2_2} shows that all the coefficients of $P_{a1}$,  $P_{a2}$, and $P_{a3}$ are positive and significant, suggesting that people are more willing to like/comment creatives from our \textbf{Prompt-for-Prompt} methods than the human-generated creatives. Also, a small $p$-value means that the differences are statistically significant. Specifically, regarding likes prompts $P_{a3}$ (social media image prompts), $P_{a1}$ (Natural social media image prompts), $P_{a2}$ (eye-catching prompts) show 2.58, 2.49, 1.85 compared to baseline (human-generated creatives), respectively. The differences between the baseline and itself are statistically significant.
In addition, regarding the comment, all 3 prompt methods achieve higher willingness than the human-generated creatives. The differences are statistically significant for $P_{a3}$ and marginally significant (i.e. p-value is between 0.05 and 0.1) for $P_{a1}$ and $P_{a2}$.


Table~\ref{tab:exp2_3} presents the effects of prompts and text-to-image methods. Midjourney and $P_{a3}$ create the best willingness to like and comment compared to the baseline (human-generated) and the difference is statistically significant (Refer to the model in Appendix A.2).


\section{Deeper Analysis}


Table~\ref{tab:tool_comparision} shows that for both the Like and Comment measure, Midjourney achieves the highest scores. For like, Midjourney records the best score from the 2 measures among 3, and the average score across all measures is 7.01, 7.79\% higher than DALL-E 3 and 17.80\% higher than Stable Diffusion. For the Comment, Midjourney shows the best score in all the 3 measures.



\begin{figure}

    \includegraphics[scale=0.45]{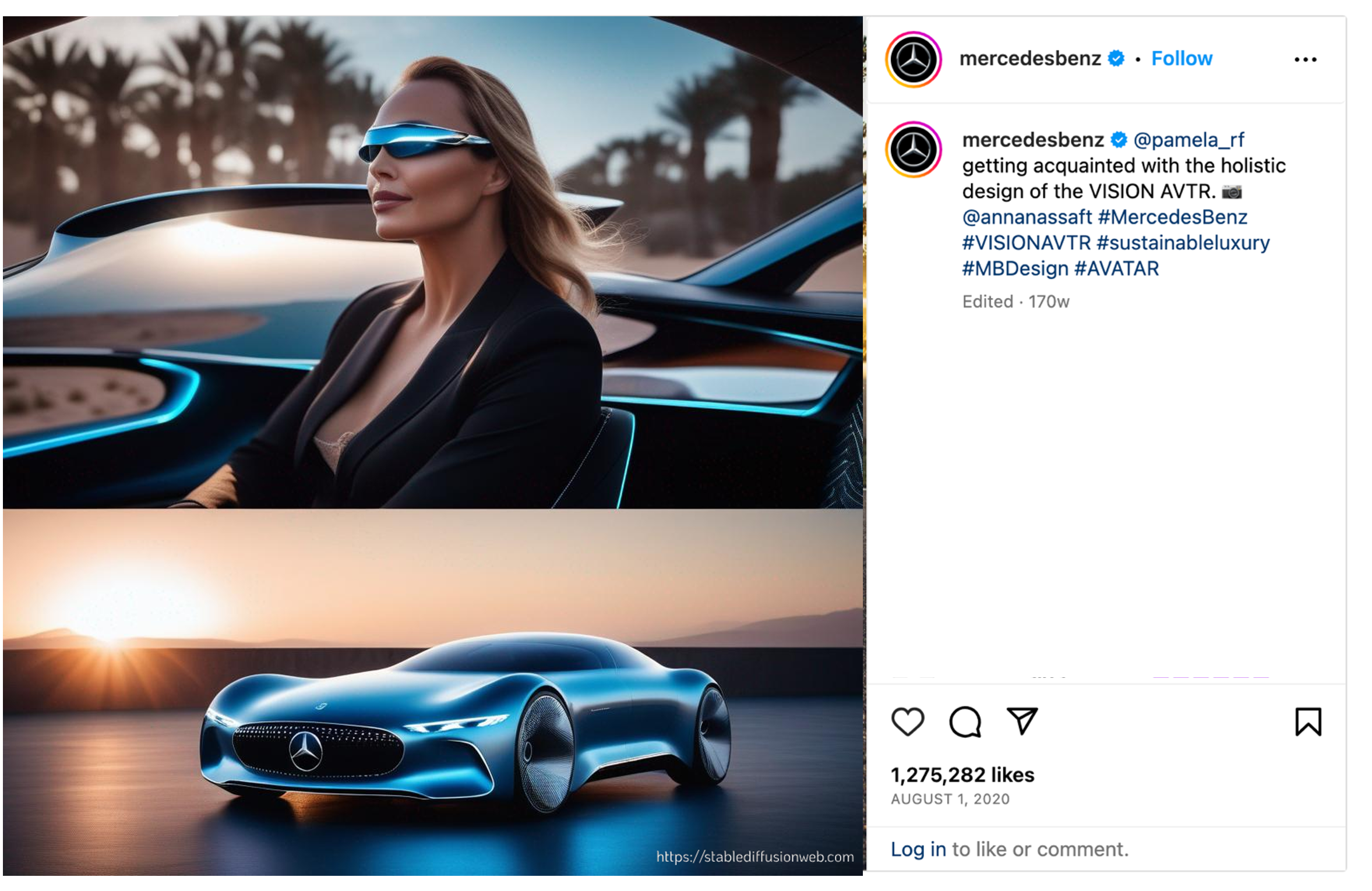}
    \caption{AI-generated creative - people+product}
    \label{fig:extra_example1}
\end{figure}

\begin{figure}
    
    \includegraphics[scale=.5]{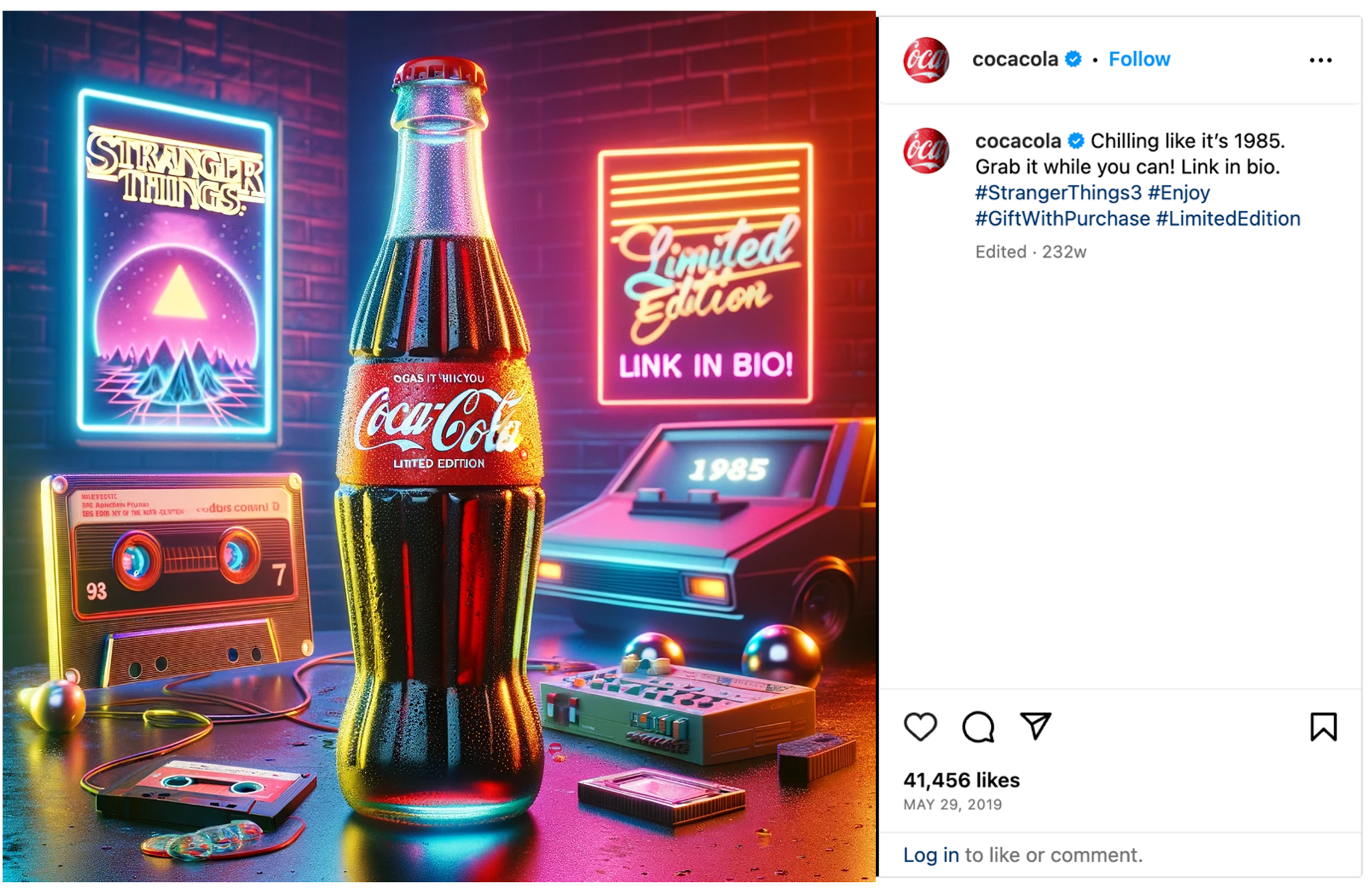}
    \caption{AI-generated creative - product-only}
    \label{fig:extra_example2}

\end{figure}

Our proposed \textbf{Prompt-for-Prompt} approach can more effectively increase social media creatives for Animal and product-only. However, they need more improvement for the creatives, including real people (Table~\ref{tab:char_img}). In detail, in animal-only creatives, AI presents improvement for 7 measures and improvement for product-only creatives from 6 measures. However, prompts fail to improve the image quality for people creatives. Prompt shows improvement only 3 measures out of 7. Figure~\ref{fig:fig1} presents an example of an animal-only image. Figure~\ref{fig:extra_example1} is a people+product image, and Figure~\ref{fig:extra_example2} is a product-only image.


We categorize our datasets into 3 groups (short, medium, and long) based on the length of the text description (Table~\ref{tab:dataset}). On average, prompts improve 20.39\% for short text descriptions compared to human-generated social media creatives while they boost 7.90\% from the medium length and 2.14\% for long text descriptions. This means that a short text description is not enough to generate high quality social media creatives, and text augmentation using \textbf{Prompt-for-Prompt} can improve
the AI's capabilities to generate social media creatives.

\section{Conclusion}
Prompt augmentation via LLMs is essential for generating creatives, even in the creative domain. However, recent research mainly covers textual data and reasoning tasks. This paper examines the effect of prompts for generating social media creatives from real Instagram data. 



\bibliographystyle{named}
\bibliography{eun}

\clearpage
   
\appendix

\setcounter{table}{0}
\setcounter{figure}{0}
\setcounter{equation}{0}

\section{Regression models}
\subsection{Augmented prompts effect}
Model~\ref{eq:reg2} is a function to show the synergy effects between text-to-image methods and augmented prompts.
\begin{align}
    Like \; or \; Comment = &\beta_{0}+\beta_{1}P_{a1}+\beta_{2}P_{a2} \nonumber \\
    +& \beta_{3}P_{a3}  \nonumber 
    + \beta_{4}Male
    + \beta_{5}Age \nonumber \\
    + & \beta_{6}Duration
    +\mu_{i}+\tau_{j} +\epsilon
    \label{eq:reg2}
\end{align}%
where : 
\begin{itemize}
    
     \item $\beta_1$, $\beta_2$, $\beta_3$ are the key coefficients of interest of this paper, which are the main effects of $P_{a1}$, $P_{a2}$, $P_{a3}$, respectively (vs human-generated)

     \item $\beta_0$ is an intercept;
     
     \item $P_{a1}$ is 1 if the creatives are generated by $P_{a1}$ -  Natural+Social Media Image or 0 otherwise;
     \item $P_{a2}$ is 1 if the creatives are generated by $P_{a2}$ - Eye-Catching+Social Media Image or 0 otherwise;
     \item $P_{a3}$ is 1 if the creatives are generated by $P_{a3}$ -Social Media Image or 0 otherwise;

    \item Male is 1 if the respondent is male or 0 otherwise;
     \item Age is an age of a participant;
     \item Duration is a time for completing human survey;
     \item $\mu_{i}$ is dummies for each brand (e.g., Apple, Coca-Cola);
     \item $\tau_{j}$ is dummies for the participants;
     \item $\epsilon$ is an error term.
\end{itemize}

\subsection{Effects of text-to-image methods and augmented prompts}
Model~\ref{eq:reg3} is a function to show the effects text-to-image methods and augmented prompts.

\begin{align}
    Like \; or \; Comment = &\beta_{0} \nonumber \\
    +& \beta_{1}MidP_{a1}+\beta_{2}MidP_{a2}+\beta_{3}MidP_{a3}\nonumber \\
+ & \beta_{4}DalP_{a1}+\beta_{5}DalP_{a2}+\beta_{6}DalP_{a3} \nonumber \\
+ & \beta_{7}SdP_{a1}+\beta_{8}SdP_{a2}+\beta_{9}SdP_{a3} \nonumber \\
+ & \beta_{10}Male + \beta_{11}Age + \beta_{12}Duration \nonumber \\
    + & \mu_{i}+\tau_{j} +\epsilon
    \label{eq:reg3}
\end{align}%
where : 
\begin{itemize}
    
     \item $\beta_1$, $\beta_2$, $\beta_3$, $\beta_4$, $\beta_5$, $\beta_6$, $\beta_7$, $\beta_8$, $\beta_9$ are the key coefficients of interest of this paper, which are the main effects of $MidP_{a1}$, $MidP_{a2}$, $MidP_{a3}$, 
     $DalP_{a1}$, $DalP_{a2}$, $DalP_{a3}$,
     $SdP_{a1}$, $SdP_{a2}$, $SdP_{a3}$, respectively (vs human-generated)

     \item $\beta_0$ is an intercept;
     
     \item $MidP_{a1}$ is 1 if the creatives are generated by Midjourney and $P_{a1}$ - Natural+Social Media Image and  or 0 otherwise;
     \item $MidP_{a2}$ is 1 if the creatives are generated by Midjourney and $P_{a2}$ - Eye-Catching+Social Media Image or 0 otherwise;
     \item $MidP_{a3}$ is 1 if the creatives are generated by Midjourney and $P_{a3}$ -Social Media Image or 0 otherwise;

     \item $DalP_{a1}$ is 1 if the creatives are generated by DALL-E 3 and $P_{a1}$ - Natural+Social Media Image and  or 0 otherwise;
     \item $DalP_{a2}$ is 1 if the creatives are generated by DALL-E 3 and $P_{a2}$ - Eye-Catching+Social Media Image or 0 otherwise;
     \item $DalP_{a3}$ is 1 if the creatives are generated by DALL-E 3 and $P_{a3}$ -Social Media Image or 0 otherwise;

     \item $SdP_{a1}$ is 1 if the creatives are generated by Stable Diffusion and $P_{a1}$ - Natural+Social Media Image and  or 0 otherwise;
     \item $SdP_{a2}$ is 1 if the creatives are generated by Stable Diffusion and $P_{a2}$ - Eye-Catching+Social Media Image or 0 otherwise;
     \item $SdP_{a3}$ is 1 if the creatives are generated by Stable Diffusion and $P_{a3}$ -Social Media Image or 0 otherwise;

    \item Male is 1 if the respondent is male or 0 otherwise;
     \item Age is an age of a participant;
     \item Duration is a time for completing human survey;
     \item $\mu_{i}$ is dummies for each brand (e.g., Apple, Coca-Cola);
     \item $\tau_{j}$ is dummies for the participants;
     \item $\epsilon$ is an error term.
\end{itemize}

\section{Examples of augmented prompt}

The full contents of augmented prompts are in Tables~\ref{tab:prompts_exp2_1_1},~\ref{tab:prompts_exp2_1}, and \ref{tab:prompts_exp2_3}.

\begin{table*}[ht!]
    \centering
    \begin{tabular}{p{0.9\linewidth} }
        \toprule
        $P_{a1}$\\ \hline

        \midrule
        
\textbf{Objective}: Design an image series that whimsically suggests a cat taking a selfie, showcasing the high-quality capabilities of the iPhone camera as captured by Rashid S.

\textbf{Scene Setup}: Position an iPhone in a scene in such a way that a cat appears to be looking into the front camera or tapping the shutter button. This could be done by placing the iPhone against a backdrop where a cat typically lounges, like a sunny windowsill, a cozy corner with scatter cushions, or an interesting outdoor locale.

\textbf{Cat Interaction}: Feature a cat interacting with the iPhone, with one paw extended towards the camera as if in the midst of taking a selfie. The cat should be the primary subject, with its personality and expression taking center stage.

\textbf{iPhone Display}: The iPhone screen should be visible, potentially reflecting the cat’s face on the screen to enhance the selfie effect. You can also show the camera app interface around the cat's reflection for a more authentic selfie look.

\textbf{Vibrant and Playful Aesthetic}: Use a bright and attractive color palette to catch the eye and evoke a playful mood. The image should use the natural light to sparkle off the cat's eyes and the iPhone screen, highlighting the quality of the photo.

\textbf{Intriguing Angles}: Experiment with various angles and perspectives to find the most engaging shot — overhead, from the side, or with the cat's face coming into the foreground and the iPhone angled towards it.

\textbf{Engagement Elements}: Add a caption or graphic element that encourages viewers to interact, such as "Guess what I'm thinking?" above the cat's selfie or a prompt for users to post their pet's selfie shot on an iPhone.

\textbf{Hashtag Integration}: Ensure that the hashtag \#ShotoniPhone is prominently featured in the image in a style that complements Apple's branding. The font should be modern and clean, matching the aesthetics of Apple's design.

\textbf{Photographer Tag}: Include a tasteful shout-out to the original photographer, @rashid\_bin\_\_, using a lower-third graphic or a discreet corner placement that doesn't take away from the overall design.

\textbf{Call to Action}: Incorporate a subtle call to action, such as “Show us your best selfie \#ShotoniPhone” to encourage user engagement and participation in the campaign.

\textbf{Brand Consistency}: Keep the image sleek and consistent with Apple's branding. The composition should be minimalistic yet striking, with a focus on the clever setup that makes the cat appear to take a selfie.

This prompt is tailored to create a playful and engaging visual story that resonates with the audience, driving home the campaign's creative angle and the technical prowess of the iPhone's camera. \\

        \bottomrule

    \end{tabular}
    \caption{Example of augmented prompt (Natural+Social Media Images) from GPT-4}
    \label{tab:prompts_exp2_1_1}
\end{table*}

\begin{table*}
    \centering
    \begin{tabular}{p{0.9\linewidth} }
        \toprule
        $P_{a1}$\\ \hline

        \midrule
        
\textbf{Objective}: Create an image that captures a cat in the act of taking a selfie, showcasing the iPhone's camera quality and the playful, creative scenario envisioned by Rashid S.

\textbf{Scene Concept}: Set up a whimsical yet plausible scenario where a cat appears to be using an iPhone to take a selfie. Consider staging the iPhone in a position that could naturally be operated by the cat, such as propped against a pile of books or resting on a soft surface within the cat's reach.

\textbf{Cat's Interaction with the iPhone}: Pose the cat with one paw extended toward the iPhone as if pressing the shutter button. Ensure the cat's face is prominently reflected on the iPhone screen, capturing a clear selfie expression.

\textbf{Background Setting}: Choose a simple yet visually appealing background that doesn't distract from the subject, such as a brightly lit room with minimal furniture or an outdoor setting with natural light.

\textbf{Props and Accessories}: Add a few playful props around the scene to enhance the storytelling element. These could include items typically associated with selfies or personalization, like sunglasses, small hats, or toys to add character to the scene.

\textbf{Photography Techniques}: Utilize a high depth of field to keep the iPhone and the cat in sharp focus while softly blurring the background. Employ lighting that flatters the cat’s features and enhances the iPhone’s screen visibility.

\textbf{Image Style and Composition}: Maintain a balance between fun and sophistication in the image composition. Align the iPhone’s branding with the creative concept to ensure the product is recognizable and the scenario is believable.

\textbf{Caption and Hashtag Integration}: Design the image with space to include an engaging caption and the hashtag \#ShotoniPhone. Ensure that the caption complements the visual content and continues the story that the image portrays.

\textbf{Photographer Tag}: Place @rashid\_bin\_\_'s handle in the image subtly but legibly to attribute the original concept and maintain connection with the campaign’s community.

\textbf{Viewer Engagement}: Encourage interaction by prompting viewers to share their interpretation of the concept or post their pet's selfies, leveraging the initial creative idea for wider campaign participation.

\textbf{Final Touches}: Apply final edits to enhance the image's clarity, color balance, and overall impact, ensuring that the image meets the high-quality standards associated with Apple's brand.

This prompt is designed to inspire the creation of an image that not only fits the playful narrative but also emphasizes the camera capabilities of the iPhone, reflecting the innovative and user-friendly reputation of Apple products.\\ \hline

        \bottomrule

    \end{tabular}
    \caption{Example of augmented prompts (Eye-catching+Social Media Images) from GPT-4}
    \label{tab:prompts_exp2_1}
\end{table*}

\begin{table*}
    \centering
    \begin{tabular}{p{0.9\linewidth} }
        \toprule
        $P_{a1}$\\ \hline

        \midrule
        
\textbf{Objective}: Craft a series of images that creatively depict a cat taking a selfie, illustrating the unique and user-centric capabilities of the iPhone's camera as visualized by photographer Rashid S.

\textbf{Scene Composition}: Set up a scene where a cat is positioned with an iPhone in such a way that it appears to be taking a selfie. This could involve the cat's paw placed on the iPhone screen as if it's pressing the shutter button or the iPhone propped up against an object with the cat looking directly into the front-facing camera.

\textbf{Cat's Pose and Expression}: Capture the cat's personality and curiosity, with its eyes focused on the screen, ensuring that the moment conveys intention, akin to a selfie pose. The cat could be in a playful state or a more serene, relaxed posture that people associate with taking selfies.

\textbf{iPhone Orientation and Visibility}: The iPhone should be visible in the frame, showing the camera interface around the cat’s reflection on the screen, adding to the selfie illusion. Make sure the iPhone model is identifiable to showcase the latest technology by Apple.

\textbf{Background and Setting}: Use a cozy and homely background, or place the cat in an unexpected but photogenic location to add an element of surprise and delight to the composition. The background should be aesthetic but not distracting, complementing the subject.

\textbf{Lighting}: Utilize natural lighting to create a soft and flattering ambiance that enhances the cat's features and the iPhone’s design. The lighting should also highlight the clarity and color accuracy of the iPhone's camera.

\textbf{Interaction with the Camera}: To emphasize the concept of the cat taking a selfie, arrange the scene to include a mirror reflection or a clear view of the cat's face in the camera's selfie mode.

\textbf{Photographer Credit}: Incorporate a credit to the contributing photographer Rashid S. (@rashid\_bin\_\_) with a discreet watermark or text overlay, ensuring it blends with the overall design and maintains the focus on the image.

\textbf{Hashtag Placement}: Place the campaign hashtag \#ShotoniPhone prominently in the image, using Apple's clean and contemporary typography for consistency with the campaign's branding.

\textbf{Viewer Engagement}: Craft the image to encourage viewers to participate in the campaign, perhaps by adding a thought bubble graphic emanating from the cat with a prompt like "Your turn to selfie" or "Can you beat my selfie game?"

\textbf{Final Touches}: Ensure the final image looks polished and vibrant, with balanced saturation and contrast to make the visual pop on social media feeds.

By following this prompt, the resulting images should engage the audience with their creativity and interactive narrative, encouraging them to view the iPhone as a device that captures not just images, but unique stories and moments.

This prompt targets the production of playful and engaging content, fitting for social media shares and resonating with the creative and high-quality essence of the Apple brand. \\ \hline

        \bottomrule

    \end{tabular}
    \caption{Example of augmented prompt (Social Media Images) from GPT-4}
    \label{tab:prompts_exp2_3}
\end{table*}

\section{Examples of AI-generated creatives from the same prompt}
\begin{figure}[ht!]
    \begin{subfigure}{\linewidth}
    \includegraphics[scale=0.45]{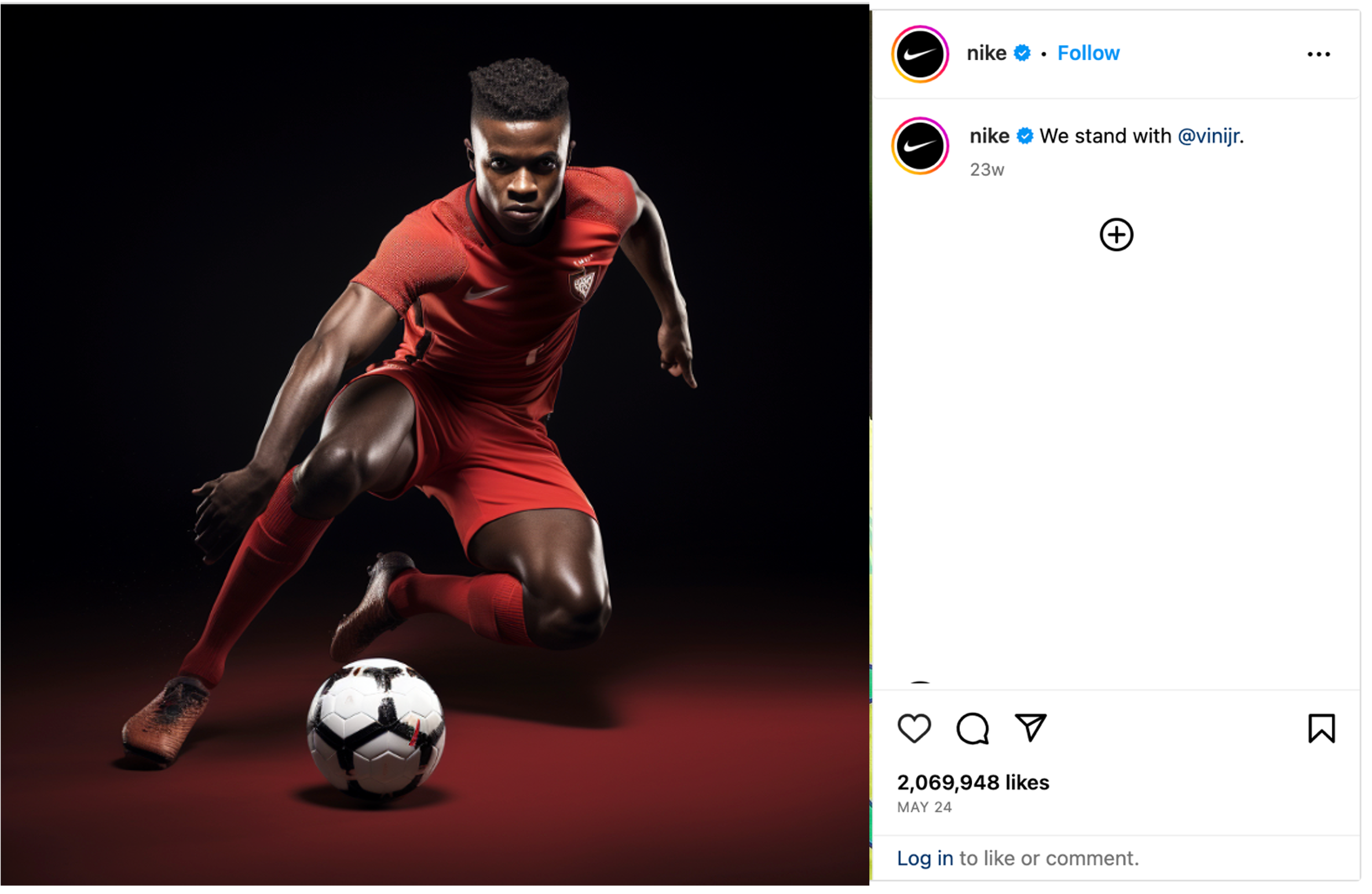}
    \caption{AI-generated image by Midjourney}
    \label{fig:extra_example1_1}
    \end{subfigure}
    
    \begin{subfigure}{\linewidth}
    \includegraphics[scale=.45]{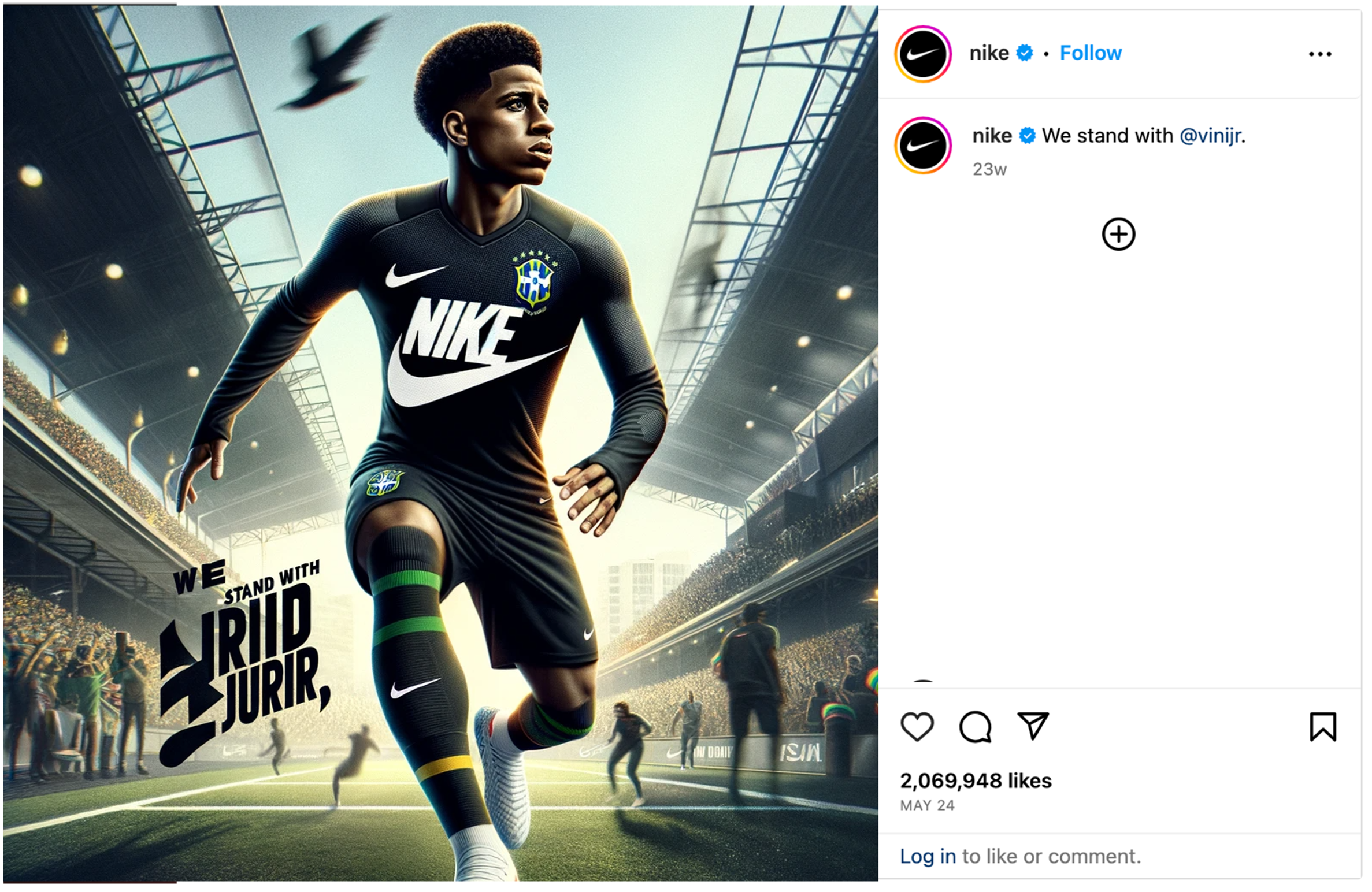}
    \caption{AI-generated image - DALL-E 3}
    \label{fig:extra_example2_1}
    \end{subfigure}

    \begin{subfigure}{\linewidth}
    \includegraphics[scale=0.45]{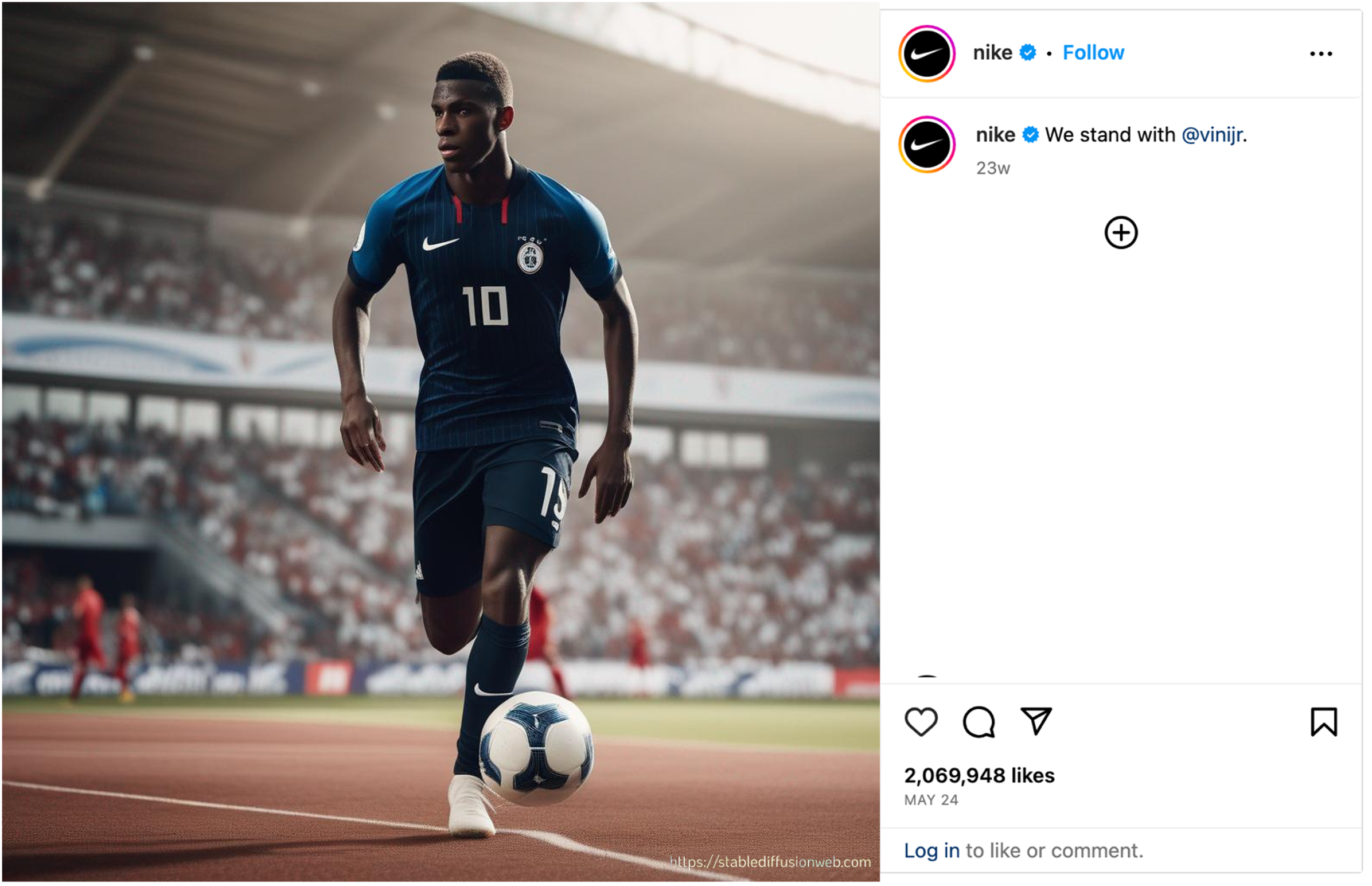}
    \caption{AI-generated image - Stable Diffusion}
    \label{fig:extra_example1_2}
    \end{subfigure}
   
    \caption{Image comparison from the same prompt}\label{fig:fig_examples1}
\end{figure}

Figure~\ref{fig:fig_examples1} illustrates a comparison of Prompt-for-Prompt outputs from the same augmented prompt (Table~\ref{tab:prompts_exp11}) using different text-to-image methods.

\begin{table*}
    \centering
    \begin{tabular}{p{0.9\linewidth}  }
        \toprule
        $P_{a1}$ \\ \hline

        \midrule
        \textbf{Composition Focus}: Create a central, striking image of Vinicius Junior (@vinijr) as the main subject, capturing his intensity on the field. Consider a freeze-frame of a memorable play or a stylized portrait that showcases his determination and spirit.

\textbf{Nike Branding}: Ensure that Vinicius is wearing Nike gear prominently displaying the Nike logo. Position the logo naturally but in a way that catches the eye, such as on his shirt or shoes.

\textbf{Color Scheme and Lighting}: Use a bold and contrasting color scheme that matches the Nike brand palette, with emphasis on stark lighting to highlight @vinijr’s features and Nike’s products. Incorporate a gradient or lighting effect that spotlights the athlete, drawing attention to the message of support.

\textbf{Support Message}: Overlay the image with the supportive slogan "We stand with @vinijr" in bold, clear typography. The font style should resonate with Nike's sleek and modern branding, and the placement should complement the image without obstructing the view of Vinicius Junior.

\textbf{Visual Effects}: Add dynamic visual effects that echo Vinicius Junior's agility and speed, such as motion blur lines trailing behind a kicked ball or speed-induced ripple effects around him.

\textbf{Background Elements}: Consider a backdrop that reflects the sport—like a blurred stadium full of fans or an abstract representation of a football pitch—to keep the focus on the athlete while still conveying the energy of the game.

\textbf{Call to Action}: Include a subtle but persuasive call to action that invites engagement from the community, such as "Show your support" or "Stand with us" with a designated hashtag that encourages user participation.

\textbf{Imagery Cohesion}: Ensure that the image looks cohesive, with all elements from Vinicius Junior to Nike branding, the supportive message, and the visual style working together to create an impactful impression.

\textbf{Authenticity}: Despite the stylization, maintain a sense of authenticity to resonate with viewers on a personal level, reminding them of Vinicius Junior’s real struggles and achievements.

\textbf{Visual Flow}: Design the image so the viewers' eyes are naturally drawn across the different elements, from Vinicius Junior to the support message and Nike branding, ensuring all components are seen.

When applying this prompt, the generated image will serve to unite Nike's audience in support of Vinicius Junior while staying true to Nike's core values and aesthetic guidelines. The result should be an engaging and supportive message wrapped in Nike's powerful brand imagery.
 \\

        \bottomrule
    \end{tabular}
    \caption{Example of augmented prompt from GPT-4}
    \label{tab:prompts_exp11}
\end{table*}

\section{Augmented prompts from other LLMs}

Table~\ref{tab:prompts_exp1_2} is an augmented prompt from Llama 2 and Table~\ref{tab:prompts_exp1} is an augmented prompt from BARD.

\begin{table*}
    \centering
    \begin{tabular}{p{0.9\linewidth}  }
        \toprule
        $P_{a1}$ \\ \hline

        \midrule
        Prompt: Create a series of social media images that showcase the unique and playful perspective of a cat taking selfies with an iPhone. The images should be natural, fun, and showcase the cat's personality while highlighting the features of the iPhone. The goal is to make the viewer feel like the cat is taking a selfie and to showcase the iPhone as the perfect tool for capturing life's moments.
        
        Image 2:A close-up shot of the cat's face as it looks directly at the camera, with the iPhone held at paw-level. The cat should have a playful expression, as if it's about to take a selfie. The background should be blurred, with the cat's fur and whiskers in sharp focus. Use the iPhone's Portrait mode to create a shallow depth of field.

        Image 2: A shot of the cat holding the iPhone and taking a selfie. The cat should be positioned at an angle, with the iPhone screen facing the viewer. The cat's paw should be visible as it taps the screen to take the photo. The background should be a colorful and fun environment, such as a cat toy or scratching post.
        
        Image 3: A shot of the cat capturing a moment with its owner. The cat and owner should be positioned in a playful pose, such as the cat sitting on the owner's lap or the owner holding the cat. The iPhone should be visible in the shot, with the cat's paw or face in the foreground. The background should be a cozy and intimate environment, such as a living room or bedroom.
        
        Image 4: A shot of the cat taking a photo of its food or water bowl. The cat should be positioned directly above the bowl, with the iPhone held at paw-level. The background should be a clean and modern kitchen or dining area. Use the iPhone's macro feature to capture the details of the food or water.Image 5: A shot of the cat taking a selfie with a fun filter or effect. The cat should be positioned directly in front of the camera, with the iPhone screen facing the viewer. The cat's paw should be visible as it taps the screen to apply the filter. The background should be a plain color or a fun pattern, such as stripes or polka dots.Image 6: A shot of the cat capturing a moment with its toys. The cat should be positioned in the center of the shot, with its toys scattered around it. The iPhone should be visible in the shot, with the cat's paw or face in the foreground. The background should be a colorful and playful environment, such as a cat playground or toy box.Image 7: A shot of the cat taking a photo of its favorite spot in the house. The cat should be positioned in its favorite spot, such as a windowsill or a cozy blanket. The iPhone should be held at paw-level, with the cat's face or paw in the foreground. The background should be a warm and inviting environment, such as a sunny window or a cozy fireplace.Image Image 8: A shot of the cat taking a selfie with its owner in the background. The cat should be positioned directly in front of the camera, with the iPhone screen facing the viewer. The cat's paw should be visible as it taps the screen to take the photo. The owner should be visible in the background, with a playful or fun expression. The background should be a colorful and fun environment, such as a park.
        
 \\

        \bottomrule
    \end{tabular}
    \caption{Example of augmented prompt from Llama 2}
    \label{tab:prompts_exp1_2}
\end{table*}

\begin{table*}
    \centering
    \begin{tabular}{p{0.9\linewidth}  }
        \toprule
        $P_{a1}$ \\ \hline

        \midrule
        1. Capture the Selfie Spirit:
        \begin{itemize}
            \item Scene: A playful cat perched on its owner's shoulder, head tilted at the camera, paw outstretched like it's holding the phone. Use a wide-angle shot to capture the playful interaction between cat and owner.
            \item Lighting: Natural light, outdoors or by a window, creating a warm and inviting atmosphere.
            \item Color Palette: Bright and engaging, emphasizing the cat's fur and eyes, complemented by the owner's clothing.
            \item Hashtag: \#ShotoniPhone \#Caturday \#PawfectSelfie
        \end{itemize}
	
2. Focus on the Details:
\begin{itemize}
    \item Close-up shot: Zoom in on the cat's face, emphasizing its expressive eyes and the way it interacts with the camera lens. Capture the slight blur of movement as it bats at the phone.
    \item Macro focus: Highlight the texture of the cat's fur, tiny whiskers, and delicate nose. Use soft lighting to create a dreamlike quality.
    \item Hashtag: \#MacroMagic \#iPhonePortrait \#SeeTheUnseen
    
\end{itemize}
	
3. Showcase the iPhone's Capabilities:
\begin{itemize}
    \item Play with depth of field: Blur the background to make the cat the main focus, highlighting its personality and independent spirit.
    \item Capture action: Use burst mode to freeze the cat mid-leap or playful swipe at the phone, emphasizing its energy and dynamism.
    \item Hashtag: \#iPhoneAction \#FastFocus \#UnleashYourInnerMeow
\end{itemize}

4. Tell a Story:

\begin{itemize}
    \item Show the behind-the-scenes: Capture the owner setting up the "selfie," adjusting the angle, and interacting with the cat to get the perfect shot.
    \item Series of images: Create a mini-storyboard showing the cat's journey from curiosity to capturing the perfect pose.
    \item Hashtag: \#StorytellingWithiPhone \#CatsofInstagram \#LifeThroughAniPhoneLens
\end{itemize}

Additional Tips:

\begin{itemize}
    \item Use props that are relatable to cats, like toys, balls, or catnip, to add visual interest and encourage interaction.
    \item Incorporate natural elements like greenery or sunlight to create a vibrant and refreshing feel.
    \item Keep the editing minimal and natural, focusing on highlighting the cat's personality and the iPhone's photographic capabilities.
By incorporating these prompts and tips, you can create natural and engaging social media images for Apple that capture the spirit of Rashid's original "ShotoniPhone" photo and connect with a wide audience of cat lovers and iPhone users.
\end{itemize}

 \\

        \bottomrule
    \end{tabular}
    \caption{Example of augmented prompt from BARD}
    \label{tab:prompts_exp1}
\end{table*}

\end{document}